\documentstyle[proceedings]{crckapb}
\input epsf
\def\msol{M_\odot}

\def\lsol{L_\odot}

\def\te{T_{eff}}
\def\wig#1{\mathrel{\hbox{\hbox to 0pt{%
\lower.5ex\hbox{$\sim$}\hss}\raise.4ex\hbox{$#1$}}}}

\begin{opening}

\title{Theory of Low Mass Stars, Brown Dwarfs and Extra-solar Giant Planets}

\author{Gilles Chabrier and Isabelle Baraffe}

\institute{C.R.A.L.,
Ecole Normale Sup\'erieure, 69364 Lyon Cedex 07, France\\
chabrier@ens-lyon.fr; ibaraffe@ens-lyon.fr}

\end{opening}
\begin{document}

\section{Introduction}
 
Accurate modeling of the mechanical and thermal properties of very-low-mass stars (VLMS), Brown Dwarfs (BD) and Extra-solar Giant Planets (EGP) is of prior importance for a wide range of physical and astrophysical problems, from the fundamental physics point of view
to the astrophysical and cosmological implications. 
They provide natural laboratories to test the different
theories,
equations of state, nuclear reaction rates, model atmospheres
aimed at describing the physics of dense and cool objects. They represent the
largest stellar population in the Galaxy, and thus provide a substantial contribution to the Galactic (disk) mass budget. Finally they represent one of the most intriguing questions in our understanding of the formation of star-like objects: are planet and star formation processes really different ? Is there, and if so what is, a minimum mass for the formation of star-like objects ?
This field has blossomed recently with the discovery of several brown dwarfs
(Nakajima et al. 1995; Rebolo et al., 1995) and numerous exoplanets since
 51 Pegasi (Mayor and Queloz 1995; Mayor, this conference), which provide important information to challenge the theory.

\section{Theoretical improvements}

VLMS, or M-dwarfs, defined hereafter as objects with
masses below 0.6$\msol$, are compact objects, with characteristic radii in the range
$\sim  0.1 - 0.6 \, R_\odot $. Their central densities and temperatures are respectively of the order
of $\rho_c\approx 10-10^3$ gcm$^{-3}$ and $T_c\approx 10^6-10^7$ K, so that the
stellar interior represents a
 strongly correlated plasma.
BD's are defined as objects not massive enough for their central temperature to sustain
hydrogen burning. This characteristic stems
from the onset of degeneracy in the contracting protostar, which prevents
an increase of the temperature as contraction proceeds. A BD therefore
never reaches thermal equilibrium and cools down for its whole life.
The hydrogen-burning minimum mass (HBMM) is $\sim 0.07 - 0.09 \msol$, depending on the initial composition from Z=Z$_{\odot}$ to Z=0, respectively (Chabrier \& Baraffe, 1997). The minimum mass depends on stellar
formation theory and is still uncertain (Hubbard, 1994).
Brown dwarfs bridge the gap in the observed mass-distribution
of astrophysical objects between the lowest-mass star
 and the largest solar planet, Jupiter
($M_J= 0.001\, \msol$). 
The distinction between BD and Giant Planets is based on different formation
scenarios, but their inner structure (apart from the central rocky core in
planets) and spectral signatures are both governed by the same physics. 

The effective temperatures of these low-mass objects are below $T_{eff}\approx 5000$ K, down to $\sim 100$ K for BDs and EGPs, and surface gravities $g=GM_\star/R_\star^2$ are
in the range $\log \,  g\approx 3.5-5.5$.
The low effective temperature 
allows the presence of stable molecules in the atmosphere (see Baraffe \& Allard, these proceedings). The presence of
these bands complicates tremendously the treatment of radiative transfer, not only because
of the numerous transitions to include in the calculations, but also because the molecular
absorption coefficients strongly depend on the frequency. Moreover molecular recombination in the interior (2H $\rightarrow$ H$_2$) leads to a decrease of the adiabatic gradient so that convection penetrates deeply into the optically-thin atmospheric layers.
Because of these physical processes, the grey-approximation is
no longer valid below $\te \sim 5000$ K, as shown by several authors (Allard, 1995; Saumon et al., 1994; Chabrier et al., 1996; Chabrier \& Baraffe, 1997).

 These conditions show that the modeling of low-mass objects requires a correct description
of non-ideal effects in the interior, as well for the equation of state (EOS) as for the screening factors of the nuclear
rates, a derivation of accurate models for dense and cool atmospheres, where molecular opacity becomes eventually
the main source of absorption, and consistent (non-grey) boundary conditions between the atmosphere and the interior along evolution.

A new EOS, devoted to the
description of low-mass stars and giant planets, has been derived recently,
which presents a consistent treatment of pressure ionization
(Saumon, Chabrier \& VanHorn 1995; SCVH), altough based on the so-called ideal-volume mixing between hydrogen and helium (see SCVH). Improvement in the field of cool atmosphere models
has blossomed within the past few years with the work of
 Allard and Hauschildt (1995; 1997), Brett (1995), and Saumon et al.
(1994). Non-grey atmosphere models now extend down to $T_{eff}$ = 900 K,
since the discovery of the first cool BD GL 299B offered a stringent
test for such models (Allard et al. 1996; Tsuji et al. 1996a; Marley et al., 1996).

A detailed description of the recent improvement in the theory of low-mass objects (LMS and BDs), and a comparison with standard grey-like treatments, is
given in a recent paper by Chabrier \& Baraffe (1997).
The modelization of VLMS, BDs and EGPs and the confrontation with observations
is examined below.

\subsection{VLM stars}

In spite of considerable progress in stellar theory - internal
structure, model atmospheres and evolution - all the VLMS models so far
failed to reproduce accurately the observed color-magnitude diagrams
(CMD) of disk or halo stars below $\sim 4000$ K, i.e. $\sim 0.4-0.6\,\msol$, 
depending on the metallicity. All the models predicted too hot an
effective temperature for a given luminosity, i.e. were too blue
compared to the observations by at least one magnitude. Such a
disagreement stemmed essentially from shortcomings both in the physics
of the interior, i.e. equation of state (EOS) and thus mass-radius
relationship and adiabatic gradient, and in the atmosphere, since all models were based on
grey atmospheres and approximate outer boundary conditions.
 Important progress has
been made recently in this field with the derivation of evolutionary models based on a {\it consistent
treatment between the interior and the atmosphere profile} (Baraffe,
Chabrier, Allard \& Hauschildt 1995, 1997; Chabrier, Baraffe \& Plez
1996). 
The effect of the outer boundary condition on the mass-$\te$ relationship has been examined in detail by Chabrier \& Baraffe (1997). These authors have made comparison with models based on different grey-like treatments and have shown
convincingly that such treatments (which imply a $T(\tau)$ relationship as an external boundary condition) are incorrect, or at best highly unreliable as soon as molecular formation sets in, i.e. for any object below $\te \wig < 5000$ K,
i.e. $m\sim 0.5\,\msol$.


The LMS models based on the afore-mentioned updated physics and consistent
(non-grey) boundary condition now reach quantitative agreement with
observations for both the disk and the halo stellar population down to the
bottom of the main sequence (Figure 1). For the {\it disk population} the models
reproduce the observed color-magnitude diagrams both in the infrared ($M_K$
vs (I-K)) and in the optical ($M_V$ vs (V-I)) (Baraffe et al., 1997b),
although below 0.1 $\msol$, $M_V\wig > 15$, the models are still too blue
by $\sim 0.2-0.4$ mag in (V-I), as illustrated in Figure 1. This is likely
to stem from the still inaccurate $TiO$ line list which shapes the flux in
the optical, or possibly from the onset of grain formation for solar
metallicity (see Baraffe \& Allard, these proceedings). More importantly,
the models are in excellent agreement with the {\it
observationally-determined} mass-magnitude relationship (Henry \& McCarthy,
1993) both in the infrared and in the optical (Chabrier, Baraffe \& Plez,
1996; Baraffe et al., 1997b) as shown in the present Figure 2 and in Figure 1 of
Baraffe \& Allard (these proceedings). For {\it metal-depleted populations}, the
models are in remarkable agreement with the main sequences of globular
clusters observed with the HST nearly down to the bottom of the main
sequence and with the halo field subdwarf sequence, as shown in Figure 1
(Baraffe et al., 1997a).  We stress that, since LMS are essentially fully
convective below $\sim 0.4\, \msol$ (Chabrier \& Baraffe, 1997), the models
are not hampered by {\it any} adjustable parameter and the agreement
between theory and observation reflects directly the reliability of the
physics entering the models. These models yield the derivation of reliable
mass-functions for the disk population (M\'era, Chabrier \& Baraffe, 1996)
and for globular clusters and halo field stars down to the brown dwarf
limit (Chabrier and M\'era, 1997).

\begin{figure}
\epsfxsize 100mm
\epsfysize=80mm
\epsfbox{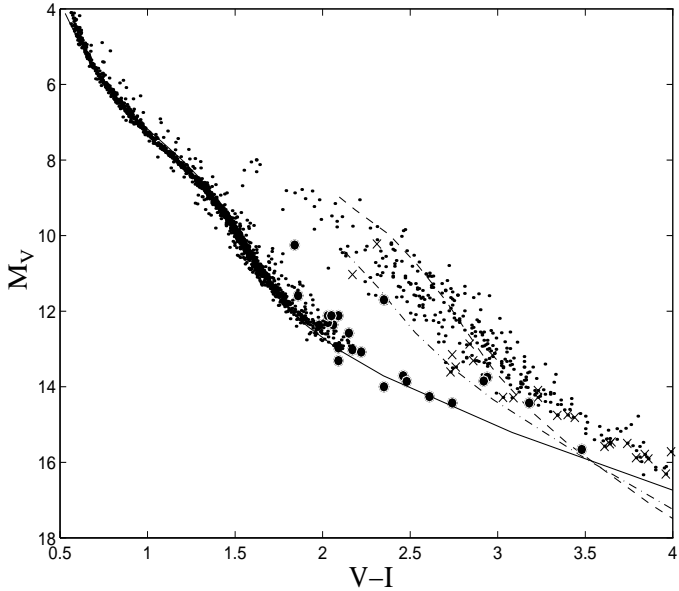}
\caption{M$_V$-(V-I) diagram for different metallicities : 
[M/H]=-1.5 (solid line), [M/H]=-0.5 (dash-dot) and
[M/H]=0 (dash). The Main Sequence of the
Globular Cluster NGC6397 from Cool et al. (1996) is shown on the left
part. Subdwarf halo field stars from Monet et al. (1992) are indicated by
full circles, as well as disk M-dwarfs of Monet et al. (1992) (crosses) and Dahn et al. (1995) (dots on the right hand side) }
\end{figure}

\begin{figure}
\epsfxsize 100mm
\epsfysize=100mm
\epsfbox{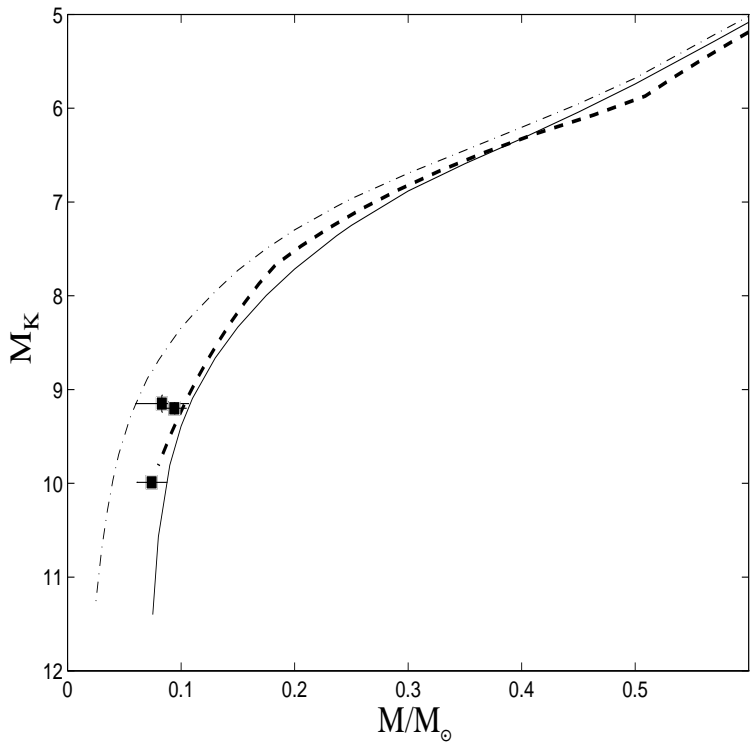}
\caption{Mass-luminosity relationship in the $K$-band. The dash-line is the fit
of the observationally-determined relation (Henry \& McCarthy, 1993; HMc93), the solid line is based on the most recent models for solar metallicity (Baraffe et al., 1997b) for $t=5$ Gyr, whereas the dash-dot line is the same models for $t=0.1$ Gyr (the ZAMS age for 0.1 $\msol$ is about 0.5 Gyr). Note that the three lowest-mass objects observed by HMc93, shown on the figure with their error bars, all exhibit strong surface (H$_\alpha$-emission) and coronal ($X$-emission) activity. Although this is not a proof of a young age, it
is a strong hint and is supported by the present analysis. }
\end{figure}

\subsection{Brown Dwarfs}

The recent discovery of substellar objects enables us to test the reliability of the theory in the BD regime. The first young brown dwarfs were
identified in the Pleiades star cluster (Rebolo et al. 1995), namely
Teide 1 and Calar 3, which both have retained their initial Lithium abundance
(Rebolo et al., 1996). Their
extremely low luminosity ($\log L/\lsol \sim -3.1$) combined with the presence
of lithium yields masses $m \sim 0.05 \msol$, well below the hydrogen-burning minimum mass (Rebolo et al., 1996; Baraffe et al, 1997b).
Given the large effective temperature of these two BDs
($\te \sim 2500 K$), due to their young age ($\sim 10^8$ yrs), the most stringent
test for extremely cool models is provided by Gl 229B (Nakajima et al. 1996).
The presence of methane in its infrared spectrum clearly reveals its substellar
nature, since CH$_4$ absorption appears below 1600~K (Tsuji et al., 1996a)
 and a star at the hydrogen burning limit has an
effective temperature of about 2000~K (Baraffe et al. 1995). Comparison of
synthetic and observed spectra yields an effective temperature of $\sim 1000 K$
(Allard et al. 1996; Tsuji et al. 1996a; Marley et al. 1996). Uncertainties in the age of the system and in the temperature of
Gl229B yield some indetermination for the mass. Evolutionary calculations based on the afore-mentioned synthetic spectra and non-grey atmosphere models yield the most likely solution $M\approx
0.04-0.055$ $\msol$ for an age $\sim 5$ Gyr, similar to our solar system (Allard et al. 1996; Marley et al., 1996).

Figure 3 presents different isochrones
as a function of metallicity (Baraffe et al. 1997b) at the bottom and below the main sequence in IR colors. The K-limit magnitudes corresponding to the HBMM
are indicated by full circles. The blue loop
displayed in the substellar domain stems from collision-induced absorption of H$_2$, and CH$_4$ absorption for solar-metallicity, in the K-band (Saumon et al., 1994; Baraffe et al., 1997a,b) which shifts
the flux back to shorter wavelengths. We predict this blueshift in IR-colors,
whereas optical colors keep reddening
almost linearly, to
be the most important photometric signature of the transition from
the stellar to the sub-stellar domain. We have made predictions of such signatures in the NICMOS filters, which should be verifiable in a very-near future (Baraffe et al., 1997a).

\begin{figure}
\epsfxsize 80mm
\epsfysize=80mm
\epsfbox{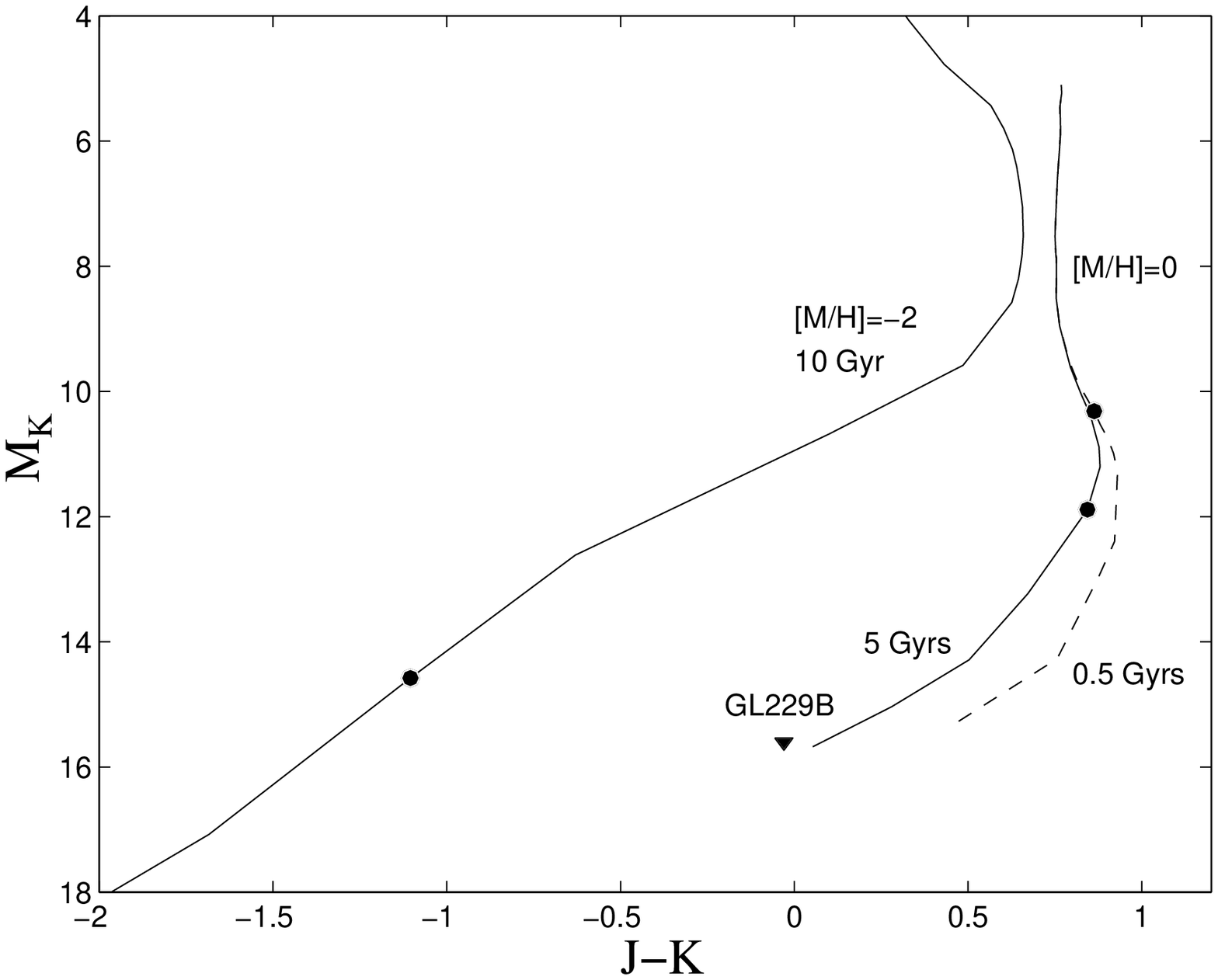}
\caption{Color-magnitude diagram in the near IR for the metallicity and ages
indicated on the figure. The full circles on the curve indicate the stellar/sub-stellar transition.}
\end{figure}

\subsection{Extrasolar Giant Planets}

A general theory of EGPs, from 0.3 to 15 Jupiter masses, and the possibility of detection by existing and future observational projects, has been performed recently by Saumon et al. (1996). These calculations take root in the theory
of solar giant planets derived previously (Chabrier et al., 1992; Guillot et al. 1995). In this first generation of calculations, the spectral
emission of the EGP and of the parent star is approximated by a black body distribution. Although 
weakened by this approximation, the predictions made in 
Saumon et al. (1996) represent the first benchmark in search strategies
for the detection of EGPs. They have been applied more specifically to the case of 51~PegA to demonstrate the
stability of giant planets orbiting nearby stars (Guillot et al., 1996).

More recent calculations by Marley et al. and Allard et al. do now predict synthetic spectra in the substellar domain from $\te \sim 2000$ K down to $\sim 300$ K. The predicted absolute fluxes of BD or EGP (Allard et al. 1997)
are displayed in Fig. 3 of Baraffe \& Allard (these proceedings)
and
compared to the sensitivity of ground and space-based observing platforms.

An extra-degree of complication in the atmosphere of objects below 
$\te \sim 2600 K$ is due to the onset of grain formation, as suggested by Tsuji and
collaborators (Tsuji et al. 1996b). This is illustrated in Figure 4 which displays the effective temperature and the radius of various exoplanets discovered recently with the domain of condensation (bars on the right hand side) of various compounds (Guillot et al., 1996).

\begin{figure}
\epsfxsize 80mm
\epsfbox{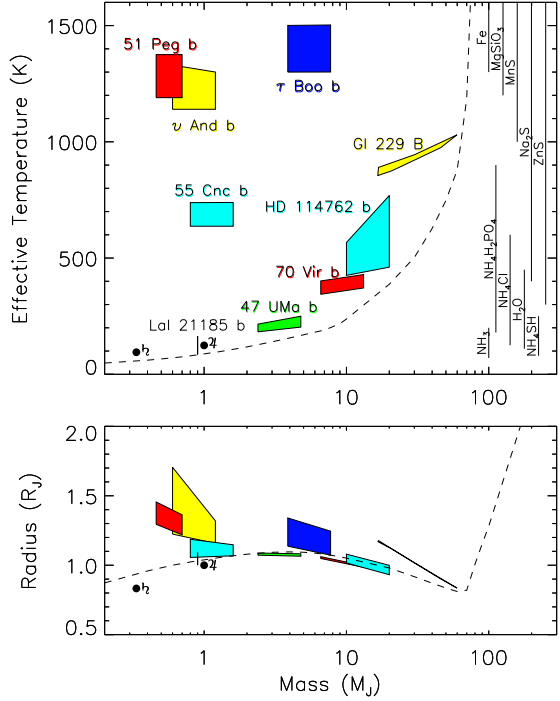}
\caption{$\te$ and radius as a function of mass for several EGPs (kindly provided by T. Guillot \& A. Burrows). Jupiter and Saturn are also indicated by their usual symbols.
The bars on the
right hand side show the domain of condensation of different species. }
\end{figure}

\section{Conclusion}

The recent improvements in the
description of the mechanical and thermal properties of cool, compact objects,
and of their photometric signature now provide solid grounds to analyse the observations and make reliable predictions. The three essential inputs in the theory of these objects are an accurate EOS, with a reliable treatment of non-ideal effects and pressure ionization, synthetic spectra with accurate molecular absorption coefficients in the optical and the IR, and consistent evolutionary calculations with correct (non-grey) boundary conditions between the atmosphere and the interior profiles.

This improved theory of LMS and substellar objects allows now the derivation of
reliable stellar mass-functions down to the brown dwarf limit and, depending on the
rate of discovery of substellar objects in the near future, will yield eventually the brown dwarf
mass function, in connection with microlensing observations.
This will bring new insight on the still unsolved problem of stellar versus planetary formation.
 
The derivation of cool atmosphere models and synthetic spectra including grain
formation represents
the next substantial improvement. Work is already under progress
in this field, both in the Tucson group and in the Lyon-Wichita group, and promises more exciting results in the near future.

%
%

\small
\tt\raggedright
\parindent=0pt
\medskip
{\bf DISCUSSION}

\medskip

PIERRE MAXTED: You suggested that two of the brown dwarfs seen in the
Pleiades are binary stars since they sit so far above the isochrones in the
HR diagram.  Is there any independent observational evidence of binarity in
these stars?

\medskip

GILLES CHABRIER: The groups of Rafael Rebolo and Gibor Basri, who suggested the possible
binarity, are examinating this possibility but, as far as I know, they don't
have a definitive answer yet.

\medskip

J\O{}RGEN CHRISTENSEN-DALSGAARD: Would it be fair to ask what the MACHO
objects are, if not brown dwarfs?

\medskip

GILLES CHABRIER: As we've shown in a recent paper (ApJ 468 L21), they can
be explained by halo white dwarfs, providing some stringent conditions are
met.  I will illustrate this point in my white dwarf talk.

\medskip

FLAVIO FUSI PECCI: You have shown nice fits of observed faint main
sequences with Galactic globulars and claim that the quality of the fits is
much better than obtained so far, thanks to the improved quality of your
models compared to previous ones.  I have seen similar fits obtained, for
instance, for NGC~6397 (HST Data) by Alexander et al.\ (1997 A\&A, 317 90)
using their own models - could you schematically explain which are the main
differences between the two sets and how you determine and evaluate the
quality of the best fitting solution?

\medskip

GILLES CHABRIER: A detailed answer is given in our papers (Chabrier \&
Baraffe 1997; Baraffe et al.\ 1997a).  To summarize, there are five important
differences: 
\begin{enumerate}

\item Our models rely on consistent non-grey boundary conditions, whereas
the Teramo models use a grey condition, based on T($\tau$) approximation. Even when modified (or adjusted), the grey
approximation is basically {\it not} correct/reliable for LMS, because of the very physics characteristic of these objects (see the present review).

\item With the {\it same physics}, and {\it no} adjustable parameter, our models reproduce all the observed sequences from Fe/H =
-2.2 (M~15) to -1.0 ($\omega$~Cen) and even -0.5 (47~Tuc), not just
NGC~6397.

\item The Teramo models assume $[M/H]=[Fe/H]$ which is not correct for metal-depleted stars, because of the $O/Fe$ enrichment, which yields $[M/H]=[O/H]$. 

\item The Teramo models rely on previous (so-called "Base") colors by Allard \& Hauschildt which overestimate the opacity (because of the Straight mean approximation, see Allard et al., 1997, ARA\&A) and are now abandoned. Moreover the bolometric corrections have been shifted arbitrarily to recover the Kurucz values at high temperatures (see their \S4).

\item Our models use consistent reddening corrections, calculated from
Allard's synthetic spectra. These corrections (E(V-I)=0.22) are in excellent
agreement with the ones used by the observers (E(V-I)=0.23; Cool et al., 1996).  The ones used by the Teramo
group (E(V-I)=0.19, see their Fig. 7) differ significantly. With the reddening and
metallicity quoted by the observers, their models are too red w.r.t. the observations (see e.g. Fig 5 of Baraffe et al., 1997a).

\end{enumerate}

\end{document}